# First-order Fragments with Successor over Infinite Words [*]

Jakub Kallas [1]    Manfred Kufleitner [2]    Alexander Lauser [2]

[1] ENS Cachan, France
[2] University of Stuttgart, FMI

**Abstract.** We consider fragments of first-order logic and as models we allow finite and infinite words simultaneously. The only binary relations apart from equality are order comparison $<$ and the successor predicate $+1$. We give characterizations of the fragments $\Sigma_2 = \Sigma_2[<,+1]$ and $\mathrm{FO}^2 = \mathrm{FO}^2[<,+1]$ in terms of algebraic and topological properties. To this end we introduce the factor topology over infinite words. It turns out that a language $L$ is in $\mathrm{FO}^2 \cap \Sigma_2$ if and only if $L$ is the interior of an $\mathrm{FO}^2$ language. Symmetrically, a language is in $\mathrm{FO}^2 \cap \Pi_2$ if and only if it is the topological closure of an $\mathrm{FO}^2$ language. The fragment $\Delta_2 = \Sigma_2 \cap \Pi_2$ contains exactly the clopen languages in $\mathrm{FO}^2$. In particular, over infinite words $\Delta_2$ is a strict subclass of $\mathrm{FO}^2$. Our characterizations yield decidability of the membership problem for all these fragments over finite and infinite words; and as a corollary we also obtain decidability for infinite words. Moreover, we give a new decidable algebraic characterization of dot-depth 3/2 over finite words.

Decidability of dot-depth 3/2 over finite words was first shown by Glaßer and Schmitz in STACS 2000, and decidability of the membership problem for $\mathrm{FO}^2$ over infinite words was shown 1998 by Wilke in his habilitation thesis whereas decidability of $\Sigma_2$ over infinite words was not known before.

*1998 ACM Subject Classification:* F.4.1 Mathematical Logic, F.4.3 Formal Languages.

## 1 Introduction

The dot-depth hierarchy of star-free languages $\mathcal{B}_n$ for $n \in \mathbb{N} + \{1/2, 1\}$ over finite words has been introduced by Brzozowski and Cohen [5]. Later, the Straubing-Thérien $\mathcal{L}_n$ hierarchy has been considered [22, 25] and a tight connection in terms of so-called wreath products was discovered [19, 23]. It is known that both hierarchies are strict [4] and that they have very natural closure properties [5, 18]. Effectively determining the level $n$ of a language in the dot-depth hierarchy or the Straubing-Thérien hierarchy is one of the most challenging open problems in automata theory. So far, the only decidable classes are $\mathcal{B}_n$ and $\mathcal{L}_n$ for $n \in \{1/2, 1, 3/2\}$, see e.g. [17] for an overview and [10] for level $\mathcal{B}_{3/2}$.

[*] The last two authors were supported by the German Research Foundation (DFG), grant DI 435/5-1.

Thomas showed that there is a one-to-one correspondence between the quantifier alternation hierarchy of first-order logic and the dot-depth hierarchy [27]. This correspondence holds if one allows $[<, +1, \min, \max]$ as a signature (we always assume that we have equality and predicates for labels of positions; in order to simplify notation, these symbols are omitted here). The same correspondence between the Straubing-Thérien hierarchy and the quantifier alternation hierarchy holds, if we restrict the signature to $[<]$, cf. [18]. In particular, all decidability results for the dot-depth hierarchy and the Straubing-Thérien hierarchy yield decidability of the membership problem for the respective levels of the quantifier alternation hierarchy.

The intersection $\Delta_2[<] = \Sigma_2[<] \cap \Pi_2[<]$ of the language classes $\Sigma_2[<]$ and $\Pi_2[<]$ of the quantifier alternation hierarchy over finite words has a huge number of different characterizations, see [24] for an overview. One of them turns out to be the first-order fragment $\mathrm{FO}^2[<]$ where one can use (and reuse) only two variables [26]. The fragment $\mathrm{FO}^2[<]$ is a natural restriction since three variables are already sufficient to express any first-order language over finite and infinite words [11]. Using the wreath product principle [23, 30], one can extend $\Delta_2[<] = \mathrm{FO}^2[<]$ to $\Delta_2[<, +1] = \mathrm{FO}^2[<, +1]$, see e.g. [14]. Decidability of $\mathrm{FO}^2[<]$ follows from the decidability of $\Sigma_2[<]$, but there is also a more direct effective characterization: A language over finite words is definable in $\mathrm{FO}^2[<]$ if and only if its syntactic monoid is in the variety **DA**, and the latter property is decidable. The wreath product principle yields **DA** $*$ **D** as an algebraic characterization of $\mathrm{FO}^2[<, +1]$, but this does not immediately help with decidability. Almeida [1] has shown that **DA** $*$ **D** $=$ **LDA**. Now, since **LDA** is decidable, membership in $\mathrm{FO}^2[<, +1]$ is decidable. Note that min and max do not yield additional expressive power for $\Delta_2[<]$ and $\mathrm{FO}^2[<]$.

Some of the characterizations and decidability results for the quantifier alternation hierarchy and for $\mathrm{FO}^2[<]$ have been extended to infinite words. Decidability of $\Sigma_1[<]$ and its Boolean closure $\mathbb{B}\Sigma_1[<]$ over infinite words is due to Perrin and Pin [15]; decidability of $\Sigma_2[<]$ over infinite words was shown by Bojańczyk [3]. The fragments $\Delta_2[<]$ and $\mathrm{FO}^2[<]$ do not coincide for infinite words. In particular, decidability of $\mathrm{FO}^2[<]$ does not follow from the respective result for $\Delta_2[<]$. Decidability of $\mathrm{FO}^2[<]$ over infinite words was first shown by Wilke [31].

Over infinite words, using a conjunction of algebraic and topological properties yields further effective characterizations of the fragments $\Sigma_2[<]$ and $\mathrm{FO}^2[<]$, cf. [7]. The key ingredient is the alphabetic topology which is a refinement of the usual Cantor topology. In addition, languages in $\mathrm{FO}^2[<] \cap \Sigma_2[<]$ can be characterized using topological notions; namely, a language $L$ over infinite words is in $\mathrm{FO}^2[<] \cap \Sigma_2[<]$ if and only if $L$ is the interior of a language in $\mathrm{FO}^2[<]$ with respect to the alphabetic topology. By complementation, a language is in $\mathrm{FO}^2[<] \cap \Pi_2[<]$ if and only if it is the topological closure of a language in $\mathrm{FO}^2[<]$. This shows that topology reveals natural properties of first-order fragments over infinite words. In this paper, we continue this line of work.

**Outline** We combine algebraic and topological properties in order to give effective characterizations of $\Sigma_2[<, +1]$ (Theorem 3.1) and $\mathrm{FO}^2[<, +1]$ (Theorem 4.1) over finite and infinite words. The key ingredient is a generalization of the alphabetic topology which we call the *factor topology*. As a byproduct, we give a new effective characterization of $\Sigma_2[<, +1]$ over finite words (Theorem 3.2), i.e., of the level 3/2 of the dot-depth hierarchy. Dually, we get a characterization of $\Pi_2[<, +1]$ over infinite words (Theorem 3.4). Moreover, we also obtain decidability results for the respective fragments over infinite words (in contrast to finite and infinite words simultaneously; Corollary 3.3 and Corollary 4.2). Concerning the intersection of fragments, we show that $L$ is in $\mathrm{FO}^2[<, +1] \cap \Sigma_2[<, +1]$ if and only if $L$ is the interior of a language in $\mathrm{FO}^2[<, +1]$ with respect to the factor topology (Theorem 6.1) and dually, $L$ is in $\mathrm{FO}^2[<, +1] \cap \Pi_2[<, +1]$ if and only if $L$ is the topological closure of a language in $\mathrm{FO}^2[<, +1]$ with respect to the factor topology (Theorem 6.2). Finally, we show that $\Delta_2[<, +1]$ is a strict subclass of $\mathrm{FO}^2[<, +1]$ and that a language $L$ is in $\Delta_2[<, +1]$ if and only if $L$ is in $\mathrm{FO}^2[<, +1]$ and clopen in the factor topology (Theorem 5.1).

## 2 Preliminaries

**Words** Throughout, $\Gamma$ is a finite alphabet and unless stated otherwise $u, v, w$ are finite words, and $\alpha, \beta, \gamma$ are finite or infinite words over the alphabet $\Gamma$. The set of all finite words is $\Gamma^*$ and the set of all infinite words is $\Gamma^\omega$. The empty word is denoted by $1$. We write $\Gamma^\infty$ for the set of all finite and infinite words $\Gamma^* \cup \Gamma^\omega$. As usual, $\Gamma^+$ is the set of all non-empty finite words $\Gamma^* \setminus \{1\}$. If $L$ is a subset of a monoid, then $L^*$ is the submonoid generated by $L$. For $L \subseteq \Gamma^*$ we let $L^\omega = \{u_1 u_2 \cdots \mid u_i \in L \text{ for all } i \geq 1\}$ be the set of infinite products. We also let $L^\infty = L^* \cup L^\omega$. The infinite product of the empty word is empty, i.e., we have $1^\omega = 1$. Thus, $L^\infty = L^\omega$ if and only if $1 \in L$. The *length* of a word $w \in \Gamma^*$ is denoted by $|w|$. We write $\Gamma^k$ for all words of length $k$ and $\Gamma^{\geq k}$ is the set of finite words of length at least $k$; similarly, $\Gamma^{<k}$ consist of all words of length less than $k$. The prefix of length $k$ of a word $w$ is denoted by $\mathrm{first}_k(w)$; it is undefined if $w$ is shorter than $k$. Symmetrically, $\mathrm{last}_k(w)$ is the suffix of $w$ of length $k$. By $\mathrm{alph}_k(\alpha)$ we denote the factors of length $k$ of $\alpha$, i.e.,

$$\mathrm{alph}_k(\alpha) = \left\{ w \in \Gamma^k \,\middle|\, \alpha = vw\beta \text{ for some } v \in \Gamma^*, \beta \in \Gamma^\infty \right\}.$$

As a special case, we have that $\mathrm{alph}_1(\alpha) = \mathrm{alph}(\alpha)$ is the *alphabet* (also called *content*) of $\alpha$. We write $\mathrm{im}_k(\alpha)$ for those factors in $\mathrm{alph}_k(\alpha)$ which have infinitely many occurrences in $\alpha$. The notation $\mathrm{im}_k(\alpha)$ comes from "*im*aginary".

**Languages** We introduce a non-standard composition $\circ$ for sufficiently long words. Let $k \geq 1$. For $u \in \Gamma^*$ and $\alpha \in \Gamma^\infty$ define $w \circ_k \alpha$ by

$$w \circ_k \alpha = vx\beta \quad \text{if there exists } x \in \Gamma^{k-1} \text{ such that } w = vx \text{ and } \alpha = x\beta.$$

Furthermore $w \circ_k 1 = w$ and $1 \circ_k \alpha = \alpha$. In all other cases $w \circ_k \alpha$ is undefined. Note that if $u \circ_k \alpha$ is defined, then $\mathrm{alph}_k(u \circ_k \alpha) = \mathrm{alph}_k(u) \cup \mathrm{alph}_k(\alpha)$. In particular, the operation $\circ_k$ does not introduce new factors of length $k$. For $A \subseteq \Gamma^k$ we define

$$\begin{aligned} A^{*_k} &= \{w_1 \circ_k \cdots \circ_k w_n \mid n \geq 0, \; w_i \in A\}, \\ A^{\omega_k} &= \{w_1 \circ_k w_2 \circ_k \cdots \mid w_i \in A\}, \\ A^{\infty_k} &= A^{*_k} \cup A^{\omega_k}, \\ A^{\mathrm{im}_k} &= \{\alpha \in \Gamma^\infty \mid \mathrm{im}_k(\alpha) = A\}. \end{aligned}$$

If $k$ is clear from the context, then we write $w \circ \alpha$ instead of $w \circ_k \alpha$, we write $A^{\circledast}$ instead of $A^{*_k}$, we write $A^{\circledcirc}$ instead of $A^{\infty_k}$, and we write $A^{\text{(im)}}$ instead of $A^{\mathrm{im}_k}$. Note that $\Gamma^* = \emptyset^{\text{(im)}}$.

A *$k$-factor monomial* is a language of the form

$$P = A_1^{\circledast} \circ u_1 \circ \cdots \circ A_s^{\circledast} \circ u_s \circ A_{s+1}^{\circledcirc}$$

for $u_i \in \Gamma^{\geq k}$ and $A_i \subseteq \Gamma^k$. The *degree* of $P$ is the length of the word $u_1 \cdots u_s$. A *$k$-factor polynomial* is a finite union of $k$-factor monomials and of words of length less than $k$. A language $L$ is a factor polynomial (resp. monomial) if there is a number $k$ such that $L$ is a $k$-factor polynomial (resp. monomial).

**Fragments of First-order Logic** We think of words as labeled linear orders, and we write $x < y$, if position $x$ comes before position $y$. Similarly, $x = y + 1$ means that $x$ is the successor of $y$. A position $x$ of a word $\alpha$ is an *a-position*, if the label of $x$ in $\alpha$ is the letter $a$.

We denote by FO the first-order logic over words. Atomic formulas in FO are $\top$ (for *true*), unary predicates $\lambda(x) = a$ for $a \in \Gamma$, and binary predicates $x < y$ and $x = y+1$ for variables $x$ and $y$. Variables range over positions in $\mathbb{N}$ and $\lambda(x) = a$ means that $x$ is an $a$-position. Formulas may be composed using Boolean connectives as well as existential quantification $\exists x \colon \varphi$ and universal quantification $\forall x \colon \varphi$ for $\varphi \in$ FO. The semantics is as usual. A *sentence* in FO is a formula without free variables. Let $\varphi \in$ FO be a sentence. We write $\alpha \models \varphi$ if $\alpha$ models $\varphi$. The *language defined by* $\varphi$ is $L(\varphi) = \{\alpha \in \Gamma^\infty \mid \alpha \models \varphi\}$.

The fragment $\Sigma_n[\mathcal{C}]$ of FO for $\mathcal{C} \subseteq \{<, +1\}$ consists of all sentences in prenex normal form with $n$ blocks of quantifiers starting with a block of existential quantifiers. In addition, only binary predicates in $\mathcal{C}$ are allowed. The fragment $\Pi_n[\mathcal{C}]$ consists of negations of formulas in $\Sigma_n[\mathcal{C}]$. We frequently identify first-order fragments with the classes of languages they define. For example, $\Delta_n[\mathcal{C}] = \Sigma_n[\mathcal{C}] \cap \Pi_n[\mathcal{C}]$ is the class of all languages which are definable in both $\Sigma_n[\mathcal{C}]$ and $\Pi_n[\mathcal{C}]$. Another important fragment is $\text{FO}^2[\mathcal{C}]$. It consists of all sentences using (and reusing) only two different names for the variables, say $x$ and $y$, and where only binary predicates from $\mathcal{C}$ are allowed. Let $\mathcal{F}$ be a fragment of first-order logic. We say that $L$ is $\mathcal{F}$-*definable over some subset* $K \subseteq \Gamma^\infty$, if there exists some formula $\varphi \in \mathcal{F}$ such that $L = \{\alpha \in K \mid \alpha \models \varphi\}$. We frequently use this notion for either $K = \Gamma^*$ or $K = \Gamma^\omega$.

**Finite Monoids** We repeat some basic notions and properties concerning finite monoids. For further details we refer to standard textbooks such as [16]. Let $M$ be a finite monoid. For every such monoid there exists a number $n \geq 1$ such that $a^n = a^{2n}$ for all $a \in M$, i.e., $a^n$ is the unique idempotent power of $a$. The set of all idempotents of $M$ is denoted by $E(M)$. We say that $M$ is *aperiodic*, if $a^n = a^{n+1}$ for all $a \in M$. If we consider a sequence $(a_1, \ldots, a_{|M|})$ of elements $a_i \in M$, then there exist $i, j \in \{1, \ldots, |M|\}$ and idempotent elements $e \in Ma_iM$ and $f \in Ma_jM$ such that $a_1 \cdots a_i e = a_1 \cdots a_i$ in $M$ and $f a_j \cdots a_{|M|} = a_j \cdots a_{|M|}$.

An important tool in the study of finite monoids are *Green's relations*. At this point, we only introduce their ordered versions $\leq_\mathcal{R}$, $\leq_\mathcal{L}$, and $\leq_\mathcal{J}$:

$$\begin{aligned} a \leq_\mathcal{R} b &\quad \Leftrightarrow \quad aM \subseteq bM, \\ a \leq_\mathcal{L} b &\quad \Leftrightarrow \quad Ma \subseteq Mb, \\ a \leq_\mathcal{J} b &\quad \Leftrightarrow \quad MaM \subseteq MbM. \end{aligned}$$

An *ordered monoid* $M$ is equipped with a partial order $\leq$ which is compatible with multiplication, i.e., $a \leq b$ and $c \leq d$ implies $ac \leq bd$. We can always assume that $M$ is ordered, since equality is a compatible partial order.

The theory of first-order fragments over finite non-empty words is presented more concisely in the context of semigroups instead of monoids. In this paper however, we want to incorporate finite and infinite words in a uniform model, and our approach is heavily based on allowing words to be empty. In order to state "semigroup conditions" for monoids, we have to use surjective homomorphisms $h \colon \Gamma^* \to M$ instead of monoids $M$ only.

Let $h \colon \Gamma^* \to M$ be a surjective homomorphism and let $e \in M$ be an idempotent. The set $P_e$ consists of all products of the form $x_0 f_1 \cdots x_{m-1} f_m x_m$ with idempotents $f_1, \ldots, f_m \in h(\Gamma^+) \subseteq M$ and elements $x_0, \ldots, x_m \in M$ satisfying the following three conditions

$$e \leq_\mathcal{R} x_0 f_1,$$
$$e \leq_\mathcal{J} f_i x_i f_{i+1} \quad \text{for all } 1 \leq i \leq m-1,$$
$$e \leq_\mathcal{L} f_m x_m.$$

If $e \notin h(\Gamma^+)$, then we set $P_e = \{1\}$. Note that in this case we necessarily have $e = 1$ in $M$. The notation $P_e$ is for *paths in $e$*. An idempotent $e$ is said to be *locally path-top* with respect to $h$ if $eP_e e \leq e$. Symmetrically, it is *locally path-bottom* with respect to $h$ if $eP_e e \geq e$. If the underlying homomorphism is clear from the context, we omit the reference to it. The homomorphism $h$ is *locally path-top* (resp. *locally path-bottom*) if all idempotents in $M$ are locally path-top (resp. locally path-bottom).

**Lemma 2.1** *Let $h : \Gamma^* \to M$ be a surjective homomorphism onto a finite monoid $M$. It is decidable whether $M$ is locally path-top.*

*Proof:* We give an algorithm computing $P_e$ for a given idempotent $e$. We define a composition on triples $T = E(M) \times M \times E(M)$ by $(f_1, x_1, f_2)(f_3, x_2, f_4) = (f_1, x_1 f_2 x_2, f_4)$ if $f_2 = f_3$. Else the composition is undefined. Compute the fixed point $P$ of the equation $P = P \cup PT_e$ with $T_e = \{(f_1, x_1, f_2) \in T \mid f_1, f_2 \in h(\Gamma^+), e \leq_\mathcal{J} f_1 x_1 f_2\}$ and initial value $P = T_e$. This requires at most $|M|^3$ iterations. Then $P_e$ is the set of all $x_0 f_1 x f_2 x_2$ where $(f_1, x, f_2) \in P$, $e \leq_\mathcal{R} x_0 f_1$ and $e \leq_\mathcal{L} f_2 x_2$. □

Let $h : \Gamma^* \to M$ be a surjective homomorphism and let $n \in \mathbb{N}$ such that $a^n$ is idempotent for all $a \in M$. Suppose that $h$ is locally path-top. With $e = a^n$, $x_0 = a$, $f_1 = a^n$, and $x_1 = 1$, we obtain $a^{n+1} = ex_0 f_1 x_1 e \leq e = a^n$ and hence,

$$a^n = a^{2n} \leq a^{2n-1} \leq \cdots \leq a^{n+1} \leq a^n$$

showing that $a^n = a^{n+1}$ for all $a \in M$, i.e., $M$ is aperiodic.

The homomorphism $h : \Gamma^* \to M$ is in **LDA** if

$$(eaebe)^n \, eae \, (eaebe)^n = (eaebe)^n$$

for all idempotents $e \in h(\Gamma^+)$ and for all $a, b \in M$. With $e = a^n$ and $b = 1$, we see that $a^{n+1} = a^n$ for all $a \in M$, i.e., $M$ is aperiodic. If the reference to the homomorphism is clear from the context, then we say "$M \in \mathcal{P}$" for some property $\mathcal{P}$ meaning that "$h \in \mathcal{P}$".

**Recognizability** A language $L \subseteq \Gamma^\infty$ is *regular* if it is recognized be some extended Büchi automaton, see e.g. [6], or equivalently, if it is definable in monadic second order logic [29]. Below, we present a more algebraic framework for recognition of $L \subseteq \Gamma^\infty$. The *syntactic preorder* $\leq_L$ over $\Gamma^*$ is defined as follows. We let $s \leq_L t$ if for all $u, v, w \in \Gamma^*$ we have the following two implications:

$$utvw^\omega \in L \Rightarrow usvw^\omega \in L \quad \text{and} \quad u(tv)^\omega \in L \Rightarrow u(sv)^\omega \in L. \tag{1}$$

Remember that $1^\omega = 1$. Two words $s, t \in \Gamma^*$ are syntactically equivalent, written as $s \equiv_L t$, if both $s \leq_L t$ and $t \leq_L s$. This is a congruence and the congruence classes $[s]_L = \{t \in \Gamma^* \mid s \equiv_L t\}$ form the *syntactic monoid* $\mathrm{Synt}(L)$ of $L$. The preorder $\leq_L$ on words induces a partial order $\leq_L$ on congruence classes, and $(\mathrm{Synt}(L), \leq_L)$ becomes an ordered monoid. It is a well-known classical result that the syntactic monoid of a regular language $L \subseteq \Gamma^\infty$ is finite, see e.g. [15, 28]. Moreover, in this case $L$ can be written as a finite union of languages of the form $[s]_L \, [t]_L^\omega$ with $s, t \in \Gamma^*$ and $st \equiv_L s$ and $t^2 \equiv_L t$.

Now, let $h : \Gamma^* \to M$ be any surjective homomorphism onto a finite ordered monoid $M$ and let $L \subseteq \Gamma^\infty$. If the reference to $h$ is clear from the context, then we denote by $[s]$ the set of finite words $h^{-1}(s)$ for $s \in M$. The following notations are used:

- $(s, e) \in M \times M$ is a *linked pair*, if $se = s$ and $e^2 = e$.

- $h$ *weakly recognizes* $L$, if
$$L = \bigcup \{[s][e]^\omega \mid (s,e) \text{ is a linked pair and } [s][e]^\omega \subseteq L\}.$$

- $h$ *strongly recognizes* $L$ (or simply *recognizes* $L$), if
$$L = \bigcup \{[s][e]^\omega \mid (s,e) \text{ is a linked pair and } [s][e]^\omega \cap L \neq \emptyset\}.$$

- $L$ is *downward closed (on finite prefixes)* for $h$, if $[s][e]^\omega \subseteq L$ implies $[t][e]^\omega \subseteq L$ for all $s, t, e \in M$ where $t \leq s$.

Using Ramsey's Theorem, one can show that for every word $\alpha \in \Gamma^\infty$ there exists a linked pair $(s, e)$ such that $\alpha \in [s][e]^\omega$. On the other hand, two different languages of the form $[s][e]^\omega$ are not necessarily disjoint. Therefore, if $L$ is weakly recognized by $h$, then there could exist some linked pair $(s, e)$ such that $[s][e]^\omega$ and $L$ are incomparable. If $L$ is strongly recognized by $h$, then for every linked pair we have either $[s][e]^\omega \subseteq L$ or $[s][e]^\omega \cap L = \emptyset$. In particular, whenever $L$ is strongly recognized by $h$, then $\Gamma^\infty \setminus L$ is also strongly recognized by $h$. Every regular language $L$ is strongly recognized by its syntactic homomorphism $h_L : \Gamma^* \to \mathrm{Synt}(L); \; s \mapsto [s]_L$. Moreover, $L$ is downward closed for $h_L$.

## 2.1 The factor topology

Topological properties play a crucial role in this paper. Very often a combination of algebraic and topological properties yields a decidable characterization of the fragments. Moreover, topology can be used to describe the relation between the fragments. This section introduces the topology matching the fragments $\Sigma_2[<, +1]$ and $\Pi_2[<, +1]$.

We define the *k-factor topology* by its basis. All sets of the form $u \circ A^\infty$ for $u \in \Gamma^*$ and $A \subseteq \Gamma^k$ are open. Therefore, singleton sets $\{u\}$ for $u \in \Gamma^*$ are open in the $k$-factor topology since $\{u\} = u \circ \emptyset^\infty$. A language is said to be *factor open* (resp. *factor closed*) if there is a natural number $k$ such that $L$ is open (resp. closed) in the $k$-factor topology.

**Proposition 2.2** *Let $L \subseteq \Gamma^\infty$ be a regular language. Then $L$ is factor open if and only if $L$ is open in the $(2\,|\mathrm{Synt}(L)|)$-factor topology.*

*Proof:* The implication from right to left is trivial. Let $n \geq 1$ be a natural number such that $L$ is open in the $n$-factor topology and let $k = 2\,|\mathrm{Synt}(L)|$. The statement is trivially true for $n \leq k$. Let $h : \Gamma^* \to \mathrm{Synt}(L)$ be a syntatic homomorphism of $L$. It strongly recognizes $L$.

In the following we shall construct for each $\alpha \in L$ a $k$-factor open environment around $\alpha$ which is contained in $L$. This is immediate if $\alpha$ is a finite word, so assume $\alpha \in \Gamma^\omega$.

For every word $x \in \Gamma^+$ of length at most $|\mathrm{Synt}(L)|$, we fix a word $f \in \Gamma^+$ of length at most $|\mathrm{Synt}(L)|$ such that $h(xf) = h(x)$ and $h(f)$ is idempotent, if such a word $f$ exists. For every word $w \in \Gamma^k$ there is a factorization $w = x_0 x_1 \hat{x}$ and $f_0, f_1 \in \Gamma^+$ such that $|x_i| \leq |\mathrm{Synt}(L)|$, $h(x_i f_i) = h(x_i)$, and $h(f_i)$ is idempotent for $i = 0, 1$. Therefore, $w' = x_0 f_0^n x_1 f_1^n \hat{x}$ has the same image under $h$ as $w$. We use for the $f_i$'s the fixed $f$'s from above.

Let $A = \mathrm{im}_k(\alpha)$ and let $\alpha^{(1)}$ be obtained from $\alpha = \alpha^{(0)}$ by replacing infinitely many occurrences of each $w \in A$ by $w'$ such that also infinitely many occurrences of each factor $w \in A$ remain unchanged. By construction, we find a common linked pair $(s, e)$ for $\alpha$ and $\alpha^{(1)}$, i.e., $\alpha, \alpha^{(1)} \in [s][e]^\omega$. Now, $\alpha \in L$ implies $[s][e]^\omega \subseteq L$ by strong recognition, and hence, $\alpha^{(1)} \in L$. We iterate this procedure of pumping idempotents and we construct $\alpha^{(i+1)}$ from $\alpha^{(i)}$ until at some point $\mathrm{im}_k(\alpha^{(i+1)}) = \mathrm{im}_k(\alpha^{(i)})$. Let $\alpha' = \alpha^{(i)}$ be the final iteration. We have $\alpha' \in L$.

Let $B = \mathrm{im}_n(\alpha')$. Since $L$ is $n$-factor open, for every sufficiently large prefix $u$ of $\alpha'$ we have $\alpha' \in u \circ B^\infty \subseteq L$. Let $C = \mathrm{im}_k(\alpha')$. We have $\alpha' \in u \circ C^\infty$ and we claim $u \circ C^\infty \subseteq L$.

Let $\beta \in u \circ C^\infty$ and let $\beta = u x_1 x_2 \cdots$ such that $|x_i| \leq |\mathrm{Synt}(L)|$ and for each $x_i$ (except maybe for the last one, if $\beta$ is finite) there exists $f_i \in \Gamma^+$ such that $h(x_i f_i) = h(x_i)$ and $h(f_i)$ is idempotent. Moreover, the $f_i$'s are in our fixed set of $f$'s from above. Consider $\beta' = u x_1 f_1^n x_2 f_2^n \cdots$ obtained from $\beta$ by "pumping idempotents". By construction of $\alpha'$, we have $\beta' \in u \circ B^\infty \subseteq L$ since every factor $f_i^n x_{i+1} f_{i+1}^n$ of $\beta'$ occurs infinitely often as a factor of $\alpha'$. By strong recognition, we see that $\beta \in L$. Let $u$ be long enough, such that when removing all pumped $f_i$'s we obtain a sufficiently large prefix $u^{(0)}$ of $\alpha$ such that $u^{(0)} x_1 x_2 \cdots \in L$, i.e., $\alpha \in u^{(0)} \circ C^\infty \subseteq L$. This shows that $L$ is $k$-factor open. □

**Proposition 2.3** *It is decidable whether a regular language $L \subseteq \Gamma^\infty$ is factor open.*

*Proof:* Lemma 2.5 below shows that for a given $k$ it is decidable whether $L$ is open in the $k$-factor topology. Proposition 2.2 gives a bound on $k$. □

**Lemma 2.4** *Let $\mathcal{A}$ be a Büchi automaton and let $L \subseteq \Gamma^\infty$ be the language accepted by $\mathcal{A}$. For any $k \geq 1$ a Büchi automaton accepting the $k$-factor interior of $L$ is effectively computable.*

*Proof:* A word $\alpha \in \Gamma^\infty$ is in the interior of $L$ if and only if there exists an open set containing $\alpha$ which is itself contained in $L$. If $\alpha$ is a finite word this is always true, so assume $\alpha \in \Gamma^\omega$. By a product automaton construction we may assume without loss that $\mathcal{A}$ always knows the last $k - 1$ symbols $a_1 \cdots a_{k-1}$ from the input. Consider a state $q$ of $\mathcal{A}$. We test whether, starting from $q$, each word in $a_1 \cdots a_{k-1} \circ A^\infty$ has an accepting computation. This is possible because the inclusion problem for Büchi automata is decidable.

Now, we modify the automaton as follows. During the computation we decide nondeterministically whether the prefix $u$ read so far is long enough and if so we guess a set of $k$-factors $A \subseteq \Gamma^k$ which we want to allow in the future such that $u \circ A^\infty$ is accepted (meaning that $A$ has passed the preceding test for the current state). With this choice we change to a new component which accepts if and only if with each new symbol $a$ we have $a_1 \ldots a_{k-1} a \in A$. If we decide that the prefix is not yet long enough, we continue in the normal computation of the original automaton. All states of the original automaton are no longer final. Therefore, a word is accepted if and only if there is a $k$-factor open subset containing the word which itself is contained in $L$. Thus the constructed automaton accepts the interior of $L$. □

**Lemma 2.5** *Let $L \subseteq \Gamma^\infty$ be a regular language and $k \geq 1$ be a natural number. It is decidable whether $L$ is open in the $k$-factor topology.*

*Proof:* A language $L$ is open if and only if it equals its interior. Using Lemma 2.4, one can construct an automaton for the interior of the language. Equivalence checking of the input automaton and the automaton for its interior is decidable, see e.g. [21]. □

## 3 The first-order fragment $\Sigma_2$

One of our main results is a decidable characterization of the fragment $\Sigma_2[<,+1]$ over finite and infinite words. It is a combination of a decidable algebraic and a decidable topological property. For finite words only, this yields a new decidable algebraic characterization for dot-depth 3/2, which in turn coincides with $\Sigma_2[<,+1]$ over finite words [27].

**Theorem 3.1** *Let $L \subseteq \Gamma^\infty$ be a regular language. The following are equivalent:*

(1) *$L$ is $\Sigma_2[<,+1]$-definable.*

(2) *$L$ is a factor polynomial.*

(3) *$L$ is factor open and there exists a surjective locally path-top homomorphism $h : \Gamma^* \to M$ which weakly recognizes $L$ such that $L$ is downward closed for $h$.*

(4) *$L$ is factor open and $\mathrm{Synt}(L)$ is locally path-top.*

The proof of the preceding theorem is given at the end of this section. Next, we give a counterpart of Theorem 3.1 for finite words, which in turn yields a new decidable characterization of dot-depth 3/2. The first decidable characterization was discovered by Glaßer and Schmitz [9, 10]. It is based on so-called forbidden patterns. Later, a decidable algebraic characterization was given by Pin and Weil [19].

**Theorem 3.2** *Let $L \subseteq \Gamma^*$ be a language. The following are equivalent over finite words:*

(1) *$L$ is $\Sigma_2[<,+1]$-definable over finite words.*

(2) *$L$ is a factor polynomial.*

(3) *$\mathrm{Synt}(L)$ is finite and locally path-top.*

*Proof:* The language $\Gamma^*$ of finite words is definable in $\Sigma_2[<]$ by stating that there is a position such that all other positions are smaller. Hence, if $L = \{w \in \Gamma^* \mid w \models \varphi\}$ for some $\varphi \in \Sigma_2[<,+1]$, then there also exists some $\varphi' \in \Sigma_2[<,+1]$ such that $L = \{\alpha \in \Gamma^\infty \mid \alpha \models \varphi'\}$. Using Theorem 3.1, this shows "1 ⇒ 2". Trivially, "2 ⇒ 3" follows from the same theorem. Finally, "3 ⇒ 1" uses the fact that every language over finite words is factor open. □

The equivalence of (1) and (2) in Theorem 3.2 was also shown by Glaßer and Schmitz using different techniques and with another formalism for defining factor polynomials [10]. As a corollary of Theorem 3.1 and Theorem 3.2 we obtain the following decidability results.

**Corollary 3.3** *Let $L$ be a regular language.*

(1) *For $L \subseteq \Gamma^\infty$ it is decidable, whether $L$ is $\Sigma_2[<,+1]$-definable.*

(2) *For $L \subseteq \Gamma^*$ it is decidable, whether $L$ is $\Sigma_2[<,+1]$-definable over finite words.*

(3) *For $L \subseteq \Gamma^\omega$ it is decidable, whether $L$ is $\Sigma_2[<,+1]$-definable over infinite words.*

*Proof:* For "1" we note that the syntactic monoid is effectively computable. Therefore, Theorem 3.1 (4) can be verified effectively by Lemma 2.1 and Proposition 2.3. Similarly, "2" follows from the decidability of Theorem 3.2 (3). The set of finite words $\Gamma^*$ is definable in $\Sigma_2[<,+1]$ over $\Gamma^\infty$. Hence, $L \subseteq \Gamma^\omega$ is $\Sigma_2[<,+1]$-definable over $\Gamma^\omega$ if and only if $L \cup \Gamma^*$ is $\Sigma_2[<,+1]$-definable over $\Gamma^\infty$, and the latter condition is decidable by "1". Therefore, assertion "3" holds. □

By duality, the properties of $\Sigma_2[<,+1]$ in Theorem 3.1 yield a decidable characterization of $\Pi_2[<,+1]$, which we state here for completeness.

**Theorem 3.4** *Let $L \subseteq \Gamma^\infty$ be a regular language. The following are equivalent:*

(1) *$L$ is $\Pi_2[<,+1]$-definable.*

(2) *$L$ is factor closed and $\mathrm{Synt}(L)$ is locally path-bottom.*

*Proof:* The language $L$ is factor closed if and only if $\Gamma^\infty \setminus L$ is factor open and moreover, the syntactic preorders of $L$ and its complement satisfy $s \leq_L t$ if and only if $t \leq_{\Gamma^\infty \setminus L} s$. Hence the claim follows by the equivalence of (1) and (4) in Theorem 3.1, since $L \in \Pi_2[<,+1]$ if and only if $\Gamma^\infty \setminus L \in \Sigma_2[<,+1]$. □

In the remainder of this section, we now prove the respective steps required for Theorem 3.1.

**Lemma 3.5** *Let $L \subseteq \Gamma^\infty$ be defined by $\varphi \in \Sigma_2[<,+1]$ and let*

$$\varphi = \exists x_1 \ldots \exists x_k \forall y_1 \ldots \forall y_k \colon \psi(x_1,\ldots,x_k,y_1,\ldots,y_k).$$

*Then $L$ is open in the $k$-factor topology.*

*Proof:* Let $\alpha \models \varphi$. We construct a $k$-open environment of $\alpha$ contained in $L$. Let $x_1,\ldots,x_k$ be such that $\psi(x_1,\ldots,x_k,y_1,\ldots,y_k)$ is true on $\alpha$ for all $y_1,\ldots,y_k$. Choose a prefix $u$ of $\alpha$ and $A \subseteq \Gamma^k$ such that $\alpha \in u \circ A^\circledast \cap A^\circledR$ and $x_i + k < |u|$ for all $i$. We claim $u \circ A^\circledast \subseteq L$. Suppose $\beta \in u \circ A^\circledast$ and $\beta \not\models \varphi$. This implies $\beta \not\models \psi(x_1,\ldots,x_k,y_1,\ldots,y_k)$ for the positions $x_i$ from above and for some positions $y_i$.

Consider $Y := \{\bar{y}_1,\ldots,\bar{y}_\ell\} \subseteq \{y_1,\ldots,y_k\}$ with $\ell$ maximal such that $\bar{y}_{i+1} = \bar{y}_i$ for $1 \leq i < \ell$, i.e., a maximal factor covered by the positions $y_i$. Take the $Y$ such that $\min Y$ is minimal. First consider the case $\bar{y}_1 \leq \max \{x_i \mid 1 \leq i \leq k\}$. Since $\ell \leq k$ we see that all positions $\bar{y}_i$ stay in the prefix $u$ and we can use the same positions in $\alpha$. If $\bar{y}_1 > \max \{x_i \mid 1 \leq i \leq k\}$. Since all factors of length $k$ appear infinitely often and $\ell \leq k$, we see that we find the factor $\beta([\bar{y}_1;\bar{y}_\ell])$ in $\alpha$ and we may choose this factor in such a way that $\bar{y}_1$ is greater than the positions of all variables already set in $\alpha$. Hence we can set the variables corresponding to those in $Y$ to the respective positions of this factor. By induction on the number of such sets $Y$, we get a distribution of the $y_i$ in $\alpha$ with the same label as the $y_i$ in $\beta$ and such that the same relations with respect to the order and successor predicate hold. Hence this distribution makes $\psi(x_1,\ldots,y_k)$ false on $\alpha$, which is a contradiction. □

**Lemma 3.6** *If $L \subseteq \Gamma^\infty$ is $\Sigma_2[<,+1]$-definable, then $\mathrm{Synt}(L)$ is locally path-top.*

*Proof:* Let $L = L(\varphi)$ with $\varphi = \exists x_1 \ldots \exists x_k \forall y_1 \ldots \forall y_k \colon \psi(x_1, \ldots, y_k)$. Consider an idempotent $e$ of $\mathrm{Synt}(L)$ and $p \in P_e$. We want to show that $epe \leq_L e$.

Consider $x_0, \ldots, x_m \in \Gamma^*$, $\bar{f}_1, \ldots, \bar{f}_m \in \Gamma^+$ such that $\bar{f}_i \equiv_L \bar{f}_i^2$ for $1 \leq i \leq m$, $e \leq_\mathcal{R} x_0 f_1$, $e \leq_\mathcal{L} f_m x_m$ and $e \leq_\mathcal{J} f_i x_i f_{i+1}$ in $\mathrm{Synt}(L)$ for $1 \leq i \leq m-1$. Let $f_i = \bar{f}_i^{k+1}$ and

$$\bar{e} = x_0 f_1 \cdot y_1 f_1 x_1 f_2 \cdots y_{m-1} f_{m-1} x_{m-1} f_m \cdot y_m f_m x_m \text{ and}$$
$$\bar{p} = x_0 f_1 x_1 \cdots f_m x_m.$$

By these properties and idempotency of $e$, we see that there exist $y_1, \ldots, y_m \in \Gamma^*$ such that $e \equiv_L \bar{e}$. Moreover, every $p \in P_e$ has such a representation, i.e., $p \equiv_L \bar{p}$. Note that $|f_i| > k$ and thus no factor of length $k$ can cover $f_i$ in total. Let

$$\alpha = u \bar{e}^{k(k+1)} \bar{p} \bar{e}^{k(k+1)} v w^\omega,$$
$$\beta = u \bar{e}^{k(k+1)} \ \bar{e}^{k(k+1)} v w^\omega.$$

We view positions of $\beta$ as a subset of the positions of $\alpha$ by omitting those positions of $\alpha$ originating from the word $\bar{p}$. Assume $\beta \models \varphi$, and let $x_i$ be such that $\psi(x_1, \ldots, y_k)$ is true on $\beta$ for all $y_j$. We claim that on $\alpha$ there is an assignment $x_i'$ such that $\psi(x_1', \ldots, y_k')$ holds for all $y_j'$.

We construct the assignment $x_i'$ by the following process. For all variables $x_i$ lying in $u$, $v$ or $w^\omega$ we set $x_i' = x_i$. Assume without restriction that the remaining variables are $x_1 < \cdots < x_\ell$. Let $X_{i,j} = \{x_i, \ldots, x_j\}$ and write $x \ll y$ whenever $y - x \geq (k+1) \cdot |\bar{e}|$ (intuitively this means that $y$ and $x$ are "far away" from each other). We start with $X_{1,\ell}$ and repeat the following until $X_{i,j}$ is empty:

- If not $x_i \ll x_{i-1}$ then we set then we set $x_i'$ so that $x_i' - x_{i-1}' = x_i - x_{i-1}$ and proceed with $X_{i+1,j}$; else

- if not $x_j \ll x_{j+1}$ then we set then we set $x_j'$ so that $x_{j+1}' - x_j' = x_{j+1} - x_j$ and proceed with $X_{i,j-1}$; else

- we have $x_i \ll x_{i-1}$ and $x_j \ll x_{j+1}$. In this case $x_i'$ is set to the position within $\bar{e}$ such that $(k+1)|\bar{e}| \leq x_{i-1}' - x_i' < (k+2)|\bar{e}|$, i.e., between $x_{i-1}'$ and $x_i'$ the factor $\bar{e}$ appears $k+1$ times. Then we proceed with $X_{i+1,j}$.

By construction, the variables $x_i'$ on $\alpha$ have the same label, relative order and successor relationship as the variables $x_i$ have on $\beta$: Although the variables may be placed in different factors $\bar{e}$, the relative position within such an factor is the same for all corresponding variables. Now, one can show that for an assignment $y_i'$ such that $\alpha \not\models \psi(x_1', \ldots, y_k')$ we find an assignment $y_i$ such that $\beta \not\models \psi(x_1, \ldots, y_k)$ contradicting the assumption. The basic idea is that, since the $f_i$ are long, all factors in $\bar{p}$ of length at most $k$ also appear in $\bar{e}$. Moreover, if a factor appears at least $k+1$ times between two variables $x_i'$ and $x_j'$ in $\alpha$ then the same holds true in $\beta$ for the variables $x_i$ and $x_j$.

Similarly, one can show $u(\bar{e}^{2k(k+1)} v)^\omega \models \varphi$ implies $u(\bar{e}^{k(k+1)} \bar{p} \bar{e}^{k(k+1)} v)^\omega \models \varphi$. In total we get $epe \equiv_L \bar{e}^{k(k+1)} \bar{p} \bar{e}^{k(k+1)} \leq_L \bar{e}^{2k(k+1)} \equiv_L e$. Since this holds for all idempotents $e$ and all $p \in P_e$ all idempotents of $\mathrm{Synt}(L)$ are locally path-top. □

The next lemma deals with the fragment $\Sigma_2[<, +1]$ over finite words, a special case which we will be needing for proving Theorem 3.1. An important tool in its proof are factorization forests. Let $M$ be a finite monoid and let $h \colon \Gamma^* \to M$ be a homomorphism. A *factorization forest* assigns to each word $w \in \Gamma^{\geq 2}$ a factorization

$$d(w) = (w_1, \ldots, w_n) \qquad \text{with } n \geq 2, \ w = w_1 \cdots w_n \text{ and } w_i \in \Gamma^+$$

such that $n \geq 3$ implies $h(w) = h(w_1) = \cdots = h(w_n)$ is idempotent. The *height* $t(w)$ of $w$ is defined by $t(a) = 0$ for leaves $a \in \Gamma$ and $t(w) = 1 + \max\{t(w_1), \ldots, t(w_n)\}$ if $d(w) = (w_1, \ldots, w_n)$. Simon's Factorization Forest Theorem [20] states that for every homomorphism $h : \Gamma^* \to M$ to a finite monoid $M$ there exists a number $t_{\max} \in \mathbb{N}$ and a factorization forest $d$ such that $t(w) \leq t_{\max}$ for all $w \in \Gamma^+$. In particular, $t_{\max}$ does not depend on $|w|$.

**Lemma 3.7** *Let $h : \Gamma^* \to M$ be a surjective homomorphism onto a finite monoid with all idempotents being locally path-top. If $h$ recognizes $L \subseteq \Gamma^*$, then $L$ is a $(2|M|)$-factor polynomial.*

*Proof:* Let $\Gamma_k = \{w \in \Gamma^* \mid k \leq |w| < 2k\}$. Now, every homomorphism $h : \Gamma^* \to M$ induces a homomorphism $h_k : \Gamma_k^* \to M$ by setting $h_k(w) = h(w)$ for all $w \in \Gamma_k$. If we apply the Factorization Forest Theorem to $h_k$, then we obtain a factorization forest for $h : \Gamma^{\geq k} \to M$ of finite height, with leaves being factors of length between $k$ and $2k$, since every word in $\Gamma^{\geq k}$ can be factored into factors in $\Gamma_k$.

Let $k = 2|M|$ and let $d$ be a factorization forest of finite height with leaves in $\Gamma_k$. By induction on the height $t(w)$ of a word $w$ of length at least $k$, we show that there exists a $k$-factor monomial $P(w)$ with degree depending only on $t(w)$ and $|M|$ such that $w \in P(w)$ and for all $u \in P(w)$ we have $h(u) \leq h(w)$. Moreover, each $P(w)$ starts and ends with a word of length at least $k$ (instead of starting and ending with a term of the form $A_i^{\circledast}$).

For leaves $w \in \Gamma_k$ we set $P(w) = w$. If $d(w) = (w_1, w_2)$, then $P(w) = P(w_1) \cdot P(w_2)$ where the dot denotes the usual concatenation. This yields a $k$-factor polynomial, since both $P(w_1)$ and $P(w_2)$ start and end with words of length at least $k$. Let now $d(w) = (w_1, \ldots, w_n)$ with $n \geq 3$ and let $e = h(w) = h(w_i)$ be the corresponding idempotent. Let $v = w_2 \cdots w_{n-1}$ be the product of the inner factors and let $A = \mathrm{alph}_k(v)$. If $|v| < 2k$, then we set $P(w) = P(w_1) \cdot v \cdot P(w_2)$. Hence, we can assume $v = sv't$ with $s, t \in \Gamma^k$. We set

$$P(w) = P(w_1) \cdot s \circ A^{\circledast} \circ t \cdot P(w_n).$$

Obviously, $w \in P(w)$. Let $u \in P(w)$ and write $u = u_1 su'tu_n$ with $u_i \in P(w_i)$ and $su't \in s \circ A^{\circledast} \circ t$. We can factorize $su't = x_0 \cdots x_m$ such that $0 < |x_i| \leq |M|$ and for each $1 \leq i \leq m$ there exists $f_i \in \Gamma^+$ such that $h(x_i) = h(f_i x_i)$ and $h(f_i)$ is idempotent. By construction of $A$, each word $x_i x_{i+1}$ is a factor of $v$ and hence

$$e \leq_{\mathcal{J}} h(x_i x_{i+1}) \leq_{\mathcal{J}} h(f_i x_i f_{i+1})$$

for each $1 \leq i < m$. Moreover, by construction of $s$ and $t$ we see that $x_0 x_1$ is a prefix of $w_2$ and that $x_m$ is a suffix of $w_{n-1}$. Together with $e = h(w_2) = h(w_{m-1})$, we obtain

$$e \leq_{\mathcal{R}} h(x_0 x_1) \leq_{\mathcal{R}} h(x_0 f_1)$$

and

$$e \leq_{\mathcal{L}} h(x_m) = h(f_m x_m).$$

By assumption, $e$ is locally path-top. Hence $e\, h(x_0 f_1 x_1 \cdots f_m x_m)\, e \leq e$ in $M$. Putting everything together yields

$$\begin{aligned}
h(u) &= h(u_1)\, h(x_0 \cdots x_m)\, h(u_n) \\
&\leq h(w_1)\, h(x_0 \cdots x_m)\, h(w_n) \\
&= e\, h(x_0 f_1 x_1 \cdots f_m x_m)\, e \leq e = h(w).
\end{aligned}$$

In all cases of the induction, the degree of $P(w)$ is bounded by $3^{t(w)} \cdot k \leq 2^{4|M|}$ where the last bound follows from the Factorization Forest Theorem for aperiodic monoids [13].

Let $L_1 = \{w \in L \mid |w| < k\}$ and let $L_2 = L \setminus L_1$. Since $L$ is recognized by $h$ we see that

$$L_2 = \bigcup_{w \in L_2} P(w)$$

and this union is finite since there are only finitely many $k$-factor monomials of degree at most $2^{4|M|}$. Therefore, $L = L_1 \cup L_2$ is a $k$-factor polynomial. $\square$

**Lemma 3.8** *Let $L \subseteq \Gamma^\infty$ be a regular language. Let $L$ be factor open and weakly recognized by a surjective locally path-top homomorphism $h : \Gamma^* \to M$ onto a finite monoid such that $L$ is downward closed on finite prefixes for $h$. Then $L$ is a factor polynomial.*

*Proof:* Let $L$ be $n$-open and let $k = \max\{2|M|, n\}$. Let $\alpha \in L$. Since $L$ is $n$-factor open, it is $k$-open. Hence, there exists $u \in \Gamma^*$ and $A = \mathrm{im}_k(\alpha)$ with $\alpha \in u \circ A^{\circledast} \subseteq L$. Since $h : \Gamma^* \to M$ is locally path-top, we know that the language $P = \{v \in \Gamma^* \mid h(v) \leq h(u)\}$ over finite words is a $k$-factor polynomial by Lemma 3.7. Moreover, we may assume that the suffix of length $k$ is explicit in all monomials of $P$. We define the factor polynomial $P_\alpha = P \circ A^{\circledast}$ and show $L = \bigcup_{\alpha \in L} P_\alpha$. Since $\alpha \in P_\alpha$ is trivial, it remains to show $P_\alpha \subseteq L$ for each $\alpha \in L$.

Let $v \in P$ and $\beta \in A^{\circledast}$ such that $v \circ \beta$ is defined. We have $u \circ \beta \in L$. Consider a linked pair $(s, e)$ with $u \circ \beta \in [s][e]^\omega \subseteq L$ and a factorization $u \circ \beta = uw\gamma$ such that $uw \in [s]$ and $\gamma \in [e]^\omega$. Let $t = h(vw)$ then $v \circ \beta = vw\gamma \in [t][e]^\omega$. Moreover $t \leq s$ and since $L$ is downward closed we have $[t][e]^\omega \subseteq L$. $\square$

**Lemma 3.9** *Let $L \subseteq \Gamma^\infty$ be a factor monomial. Then $L$ is definable in $\Sigma_2[<, +1]$.*

*Proof:* Let $L = A_1^{\circledast} \circ u_1 \circ \cdots \circ A_s^{\circledast} \circ u_s \circ A_{s+1}^{\circledcirc}$ for $A_i \subseteq \Gamma^k$, $u_i \in \Gamma^{\geq k}$. Without restriction we assume $|u_i| = k$. Consider the formula

$$\exists x_1 \ldots \exists x_s \forall y \colon \Big( \bigwedge_{1 \leq i \leq s} \lambda(x_i) = u_i \wedge x_i < x_{i+1} \Big) \wedge \bigwedge_{1 \leq i \leq s+1} \psi_i. \tag{2}$$

The first conjunction states that for each $i$, $x_i$ is the position of the marker $u_i$, and that the markers appear in the correct order. The formula $\psi_i$ imposes the factor alphabetic restriction $A_i$ between $u_{i-1}$ and $u_i$. More precisely, $\psi_i$ is set to $x_{i-1} < y < x_i \Rightarrow \lambda(y) \in A_i$. In these formulas, we use the conventions $x_0 = 0$ and $x_{s+1} = \infty$; the expression $\lambda(x_i) = u_i$ (resp. $\lambda(y) \in A_i$) is an abbreviation saying that at position $x_i$ the factor $u_i$ begins (resp. at position $y$ some factor in $A_i$ begins). These abbreviations are readily replaced in such a way that the formula remains in $\Sigma_2[<, +1]$. Therefore, $L$ is defined by the $\Sigma_2[<, +1]$-formula given in (2). $\square$

We conclude this section with the proof of Theorem 3.1.

*Proof (Theorem 3.1):* "$1 \Rightarrow 4$": This is Lemma 3.5 together with Lemma 3.6.

"$4 \Rightarrow 3$": Strong recognition implies weak recognition. The claim follows because the syntactic homomorphism $h_L : \Gamma^* \to \mathrm{Synt}(L)$ strongly recognizes $L$.

"$3 \Rightarrow 2$": This follows from Lemma 3.8.

"$2 \Rightarrow 1$": Let $L$ be a union of factor monomials and a (finite) set $K$ of words of length less than $k$. By Lemma 3.9 each monomial is definable in $\Sigma_2[<, +1]$ and of course so is $K$. The result follows since $\Sigma_2[<, +1]$ is closed under union. $\square$

## 4 First-order logic with two variables

In this section, we consider two-variable first-order logic with order and successor predicates $[<, +1]$ over finite and infinite words. The fragment $\mathrm{FO}^2[<, +1]$ admits a temporal logic counterpart having the same expressive power [8]. It is based on unary modalities only. Wilke [31] has shown that membership is decidable for $\mathrm{FO}^2[<, +1]$. We complement these results by giving a simple algebraic characterization of this fragment. An important concept in our proof is a refinement of the factor topology. A set of the form $A^{\text{\tiny ⓜ}}$ is definable in $\mathrm{FO}^2[<, +1]$ but it is neither open nor closed in the factor topology. This observation leads to the *strict $k$-factor topology*. A basis of this topology is given by all sets of the form $u \circ A^{\infty} \cap A^{\text{\tiny ⓜ}}$ for $u \in \Gamma^*$ and $A \subseteq \Gamma^k$. We do not use this topology outside this section. Using the refined topology and the class **LDA** we can now state the following theorem.

**Theorem 4.1** *Let $L \subseteq \Gamma^{\infty}$ be a regular language. The following are equivalent:*

(1) *$L$ is $\mathrm{FO}^2[<, +1]$-definable.*

(2) *$L$ is weakly recognized by some homomorphism $h : \Gamma^* \to M \in \mathbf{LDA}$ and closed in the strict $(2\,|M|)$-factor topology.*

(3) $\mathrm{Synt}(L) \in \mathbf{LDA}$.

The proof of the above theorem can be found at the end of this section. The syntactic monoid of a regular language is effectively computable. Hence, one can verify whether property (3) in Theorem 4.1 holds. Since both $\Gamma^*$ and $\Gamma^{\omega}$ are $\mathrm{FO}^2[<, +1]$-definable over $\Gamma^{\infty}$, this immediately gives us the following corollary.

**Corollary 4.2** *Let $L$ be a regular language.*

(1) *For $L \subseteq \Gamma^{\infty}$ it is decidable, whether $L$ is $\mathrm{FO}^2[<, +1]$-definable.*

(2) *For $L \subseteq \Gamma^*$ it is decidable, whether $L$ is $\mathrm{FO}^2[<, +1]$-definable over finite words.*

(3) *For $L \subseteq \Gamma^{\omega}$ it is decidable, whether $L$ is $\mathrm{FO}^2[<, +1]$-definable over infinite words.* □

The following proposition relates monoids in **LDA** with monoids which are simultaneously locally path-top and locally path-bottom. It is a useful tool in the proof of Theorem 4.1. Moreover, it immediately follows that $\Delta_2[<, +1]$ is a subset of $\mathrm{FO}^2[<, +1]$. We will further explore the relation between these two fragments in the next section.

**Proposition 4.3** *Let $M$ be finite and let $h : \Gamma^* \to M$ be a homomorphism of monoids. The following are equivalent:*

(1) $M \in \mathbf{LDA}$.

(2) $eP_e e = e$ *for all idempotents $e$ of $M$.*

*Proof:* "1 ⇒ 2" Let $n \in \mathbb{N}$ with $a^n = a^{2n}$ for all $a \in M$. First suppose $m = 1$. In particular, $f_1 = f_m$ and $x_1 = x_m$. Let $e = x_0 f_1 b = c f_1 x_1$ for some $b, c \in M$. Then

$$\begin{aligned} e &= (x_0 f_1 b\, c f_1 x_1)^n x_0 f_1 b \\ &= x_0 (f_1 b c f_1 x_1 x_0 f_1)^n b \\ &= x_0 (f_1 b c f_1 x_1 x_0 f_1)^n\, f_1 x_1 x_0 f_1\, (f_1 b c f_1 x_1 x_0 f_1)^n b \quad \text{by } h \in \mathbf{LDA} \\ &= (x_0 f_1 b\, c f_1 x_1)^n\, x_0 f_1 x_1\, (x_0 f_1 b\, c f_1 x_1)^n x_0 f_1 b \\ &= e x_0 f_1 x_1 e. \end{aligned}$$

Let now $m > 1$ and let $e = b f_1 x_1 f_2 c$ for some $b, c \in M$. Set $x_1' = b f_1 x_1$. By induction we see that

$$e = e x_1' f_2 \cdots x_{m-1} f_m x_m e.$$

Set $x_1'' = x_1 f_2 \cdots x_{m-1} f_m x_m e$. From the case $m = 1$ we obtain

$$e = e x_0 f_1 x_1'' e = e x_0 f_1 x_1 \cdots f_m x_m e.$$

Note that indeed $e \leq_\mathcal{R} x_1' f_2$ and $e \leq_\mathcal{L} f_1 x_1''$.

"2 ⇒ 1" Let $e \in h(\Gamma^+) \subseteq M$ be idempotent and $x, y \in M$. Setting $g = (exeye)^\omega$, $x_0 = f_1 = e = f_2 = x_2$, $x_1 = x$ we see that $x_0 f_1 x_1 f_2 x_2 = exe \in P_g$. Therefore, $(exeye)^\omega exe(exeye)^\omega = gexeg = g = (exeye)^\omega$ and hence $M \in \mathbf{LDA}$. □

**Example 4.4** Let $\Gamma = \{a, b, c\}$. Consider the language $L_1 = \Gamma^* a b^* a \Gamma^\infty$ consisting of all words such that there are two $a$'s that only contain $b$'s in between. It is easy to see that $L_1$ is $\Sigma_2[<]$-definable. Next, we will show that $L_1$ is not $\mathrm{FO}^2[<, +1]$-definable. Choose $n \in \mathbb{N}$ such that $s^n$ is idempotent for every $s \in \mathrm{Synt}(L_1)$. Then

$$(b^n a b^n c b^n)^n \notin L_1 \quad \text{whereas} \quad (b^n a b^n c b^n)^n b^n a b^n (b^n a b^n c b^n)^n \in L_1.$$

This shows that $\mathrm{Synt}(L_1)$ is not in $\mathbf{LDA}$. By Theorem 4.1 we conclude that $L_1$ is not $\mathrm{FO}^2[<, +1]$-definable. Similarly, $L_2 = \Gamma^\infty \setminus L_1$ is definable in $\Pi_2[<]$ but not in $\mathrm{FO}^2[<, +1]$. ◇

**Lemma 4.5** *Let $L \subseteq \Gamma^\infty$ be definable in $\mathrm{FO}^2[<, +1]$. Then $\mathrm{Synt}(L) \in \mathbf{LDA}$.*

*Proof:* Let $L$ be defined by a $\mathrm{FO}^2[<, +1]$-formula of quantifier depth $m$. Choose $\bar{e} \in \Gamma^+$, $s, t \in \Gamma^*$ and $n > m$ such that all $n$-powers are idempotent in $\mathrm{Synt}(L)$. Let $e = \bar{e}^n$. Note that $|e| > m$, i.e., no factor of length at most $m$ can cover the whole factor $e$. We show in the following that $(esete)^n ese(esete)^n \equiv_L (esete)^{2n}$.

Let $u, v, w \in \Gamma^*$ and $\alpha = u(esete)^n ese(esete)^n vw^\omega$ and $\beta = u(esete)^n (esete)^n vw^\omega$. We identify the positions of $\beta$ with a subset of the positions in $\alpha$ in the natural way. Note that in particular the successor of the last position in the prefix $u(esete)^n$ of $\beta$ is the first position of the suffix $(esete)^n vw^\omega$. We use $x, y$ to designate positions of $\alpha$ and $x', y'$ for positions of $\beta$.

We define balls $B_i$ around the difference of $\alpha$ and $\beta$ in the following way:

$$\alpha = u(esete)^{n-i-1} \underbrace{esete \overbrace{(esete)^i ese(esete)^i}^{B_i} esete}_{B_{i+1}} (esete)^{n-i-1} vw^\omega.$$

Therefore, the set of positions of $\beta$ are all positions of $\alpha$ except those that are in $B_0$.

The $i$-context $\lambda_i(z)$ of a position $z$ on a word $\gamma$ is the factor induced by the positions $[z-i; z+1]$ (which may be shorter than $2i+1$ if $z$ lies near the boundary of $\gamma$). We say that a tuple $(x, y, x', y')$ is $i$-*legal* if

$$x', y' \notin B_0,$$
$$x = y \text{ iff } x' = y',$$
$$x = y \pm 1 \text{ iff } x' = y' \pm 1,$$
$$x < y \text{ iff } x' < y',$$
$$\lambda_{m-i}(x) = \lambda_{m-i}(x') \text{ and } \lambda_{m-i}(y) = \lambda_{m-i}(y').$$

The idea is that $x'$, $y'$ are positions in $\beta$ and the configuration cannot be distinguished by atomic formulas and checking the contexts of the positions up to width $m - i$. We say that $(x, y, x', y')$ is $i$-*close* if it is $i$-legal and

$$x \neq x' \Rightarrow x, x' \in B_i,$$
$$y \neq y' \Rightarrow y, y' \in B_i,$$

that is, in addition to being legal, the respective positions either are the same or if they are not the same then they are both "not to far" from $B_0$. So if either $z$ or $z'$ is not in $B_i$ we can deduce that $z = z'$.

At the beginning we have $x = y = x' = y'$ are the first position in $\alpha$ and $\beta$ and this configuration is 0-close since $x'$ and $y'$ cannot be in $B_0$ because $e$ is longer than $m$.

Now we claim that if $(x, y, x', y')$ is $i$-close and $\varphi(x, y) \in \text{FO}^2[<, +1]$ has quantifier depth $\leq m - i$ than

$$\alpha, x, y \models \varphi(x, y) \quad \Leftrightarrow \quad \beta, x', y' \models \varphi(x', y').$$

For $i = n$ this is immediate due to the fact that the situation is 0-legal, all atomic formulas agree on their value on $\alpha$ and $\beta$. Let now $i < n$. We may assume without loss that $\varphi(x, y) = \exists x \colon \psi(x, y)$. Let $\alpha, x, y \models \varphi(x, y)$. Then there is $\tilde{x}$ such that $\psi(\tilde{x}, y)$ is true on $\alpha$. First consider the case $\tilde{x} = y$. We set $\tilde{x}' = y'$ and see that $(\tilde{x}, y, \tilde{x}', y')$ is $(i + 1)$-close. Note that here we use that $e$ is long enough so that a context "near the middle" cannot extend into the $s$ in $B_0$. Hence by induction $\psi(\tilde{x}', y')$ is true on $\beta$ and therefore $\beta, x', y' \models \varphi(x', y')$.

Consider now the case $\tilde{x} = y \pm 1$. Then we set $\tilde{x}' = y' \pm 1$. This situation is again $(i + 1)$-close (here we use that in $\beta$ the successor of the last position before $B_0$ is the first position after $B_0$).

Now consider $\tilde{x} + 1 < y$. If $\tilde{x} \notin B_i$ then we set $\tilde{x}' = \tilde{x}$. In this situation we have $\tilde{x}' + 1 \leq y'$. Moreover we have equality only if $y'$ is the first position in $B_i$. Hence, by choice of $e$, we find a position $\hat{x}'$ in $B_{i+1}$ with the same $m - (i + 1)$-context such that $\hat{x}' + 1 < y'$ and we obtain an $(i + 1)$-close situation for both cases. If $\tilde{x} + 1 < y$ and $\tilde{x} \in B_i$ we choose $\tilde{x}' \in B_{i+1} \setminus B_i$ to the first position with the same $m - (i + 1)$ context. This is possible, again by choice of $e$. We handle $\tilde{x} > y + 1$ similarly. We showed that starting with an $i$-close situation, we always can assure an $(i + 1)$-close situation for $\psi(x, y)$ with quantifier depth $\leq m - i - 1$. By induction this shows that $\alpha, x, y \models \varphi(x, y)$ implies $\beta, x', y' \models \varphi(x', y')$. The reverse implication is obtained by a symmetric argumentation.

Taking $i = 0$ in the claim above the first requirement in (1) follows. By similar arguments we see that formulas of quantifier depth at most $m$ agree on the words $u\big((esete)^n ese(esete)^n v\big)^\omega$ and $u\big((esete)^n (esete)^n v\big)^\omega$. This shows that $h_L \colon \Gamma^* \to \text{Synt}(L)$ is in **LDA**. $\square$

**Lemma 4.6** *If $L \subseteq \Gamma^\infty$ is recognized by $h \colon \Gamma^* \to M$ in **LDA**, then $L$ is clopen in the strict $k$-factor topology for every $k \geq 2\lvert M \rvert$.*

*Proof:* Since $\Gamma^\infty \setminus L$ is also recognized by $h$, it suffices to show that $L$ is open. Let $\alpha \in [s][e]^\omega \subseteq L$ for some linked pair $(s,e) \in M^2$ and let $A = \mathrm{im}_k(\alpha) \neq \emptyset$. We write $\alpha = s_0 e_1 e_2 \cdots$ with $h(s_0) = s$, $h(e_i) = e$, and $e_1 e_2 \cdots \in A^\circledast$. Moreover, we can assume $|e_i| \geq k$ and $\alpha_k(e_i) = A$ for each $i \geq 1$. Let $r_1$ be the prefix of $e_1$ of length $k-1$. We have $\alpha \in s_0 r_1 \circ A^\circledast \cap A^{\circledast_{\mathrm{im}}}$.

We show that $s_0 r_1 \circ A^\circledast \cap A^{\circledast_{\mathrm{im}}} \subseteq L$ which proves the claim. Let $\beta \in s_0 r_1 \circ A^\circledast \cap A^{\circledast_{\mathrm{im}}}$ and write $\beta = s_0 r_1 r_2 f_1 f_2 \cdots$ such that $f = h(f_1) = h(f_2) = \ldots$ and $\bigl(h(r_1 r_2), f\bigr)$ is a linked pair with $\mathrm{alph}_k(f_i) = A$ for all $i \geq 1$. Let $r = h(r_1 r_2)$.

We factorize $r_1 r_2 f_1 = x_0 x_1 \cdots x_m$ such that $|x_i| \leq |M|$ and for each $x_i$, $i < m$ there exists an idempotent $g_{i+1} \in h(\Gamma^+) \subseteq M$ with $h(x_i) g_{i+1} = h(x_i)$. By construction of $k$ and $r_1$ we see that $x_0$ is a prefix of $r_1$. Hence,
$$e \leq_\mathcal{R} h(r_1) \leq_\mathcal{R} h(x_0) = h(x_0)\, g_1.$$

By choice of $A$ and $e_1$, we see that for all $0 < i \leq m$, the word $x_{i-1} x_i$ is a factor of $e_1$. Hence, for all $1 < i < m$ we have
$$e \leq_\mathcal{J} h(x_{i-1} x_i) = h(x_{i-1})\, g_i\, h(x_i)\, g_{i+1} \leq_\mathcal{J} g_i\, h(x_i)\, g_{i+1}.$$

Since $x_{m-1} x_m$ is a factor of $e_1$, there exists $t_0 \in \Gamma^*$ such that $x_{m-1} x_m t_0$ is a suffix of $e_1$. With $t = h(t_0)$ we see that
$$e \leq_\mathcal{L} h(x_{m-1} x_m) t = h(x_{m-1}) g_m h(x_m) t \leq_\mathcal{L} g_m h(x_m) t.$$

By Proposition 4.3 we see that
$$e = e h(x_0) g_1 h(x_1) \cdots g_m h(x_m) t e = e h(r_1 r_2 f_1) t e = e r f t e$$

Similarly, using $\mathrm{alph}_k(f_i) = A$, we obtain $p, q \in M$ with $f = f p e q f$. Since $M$ is aperiodic, there exists $n \in \mathbb{N}$ such that $a^n = a^{n+1}$ for all $a \in M$. It follows
$$e = e r f p e q f t e = (e r f p)^n\, e\, (q f t e)^n = (e r f p)^{n+1}\, e\, (q f t e)^n = e r f p e$$

and similarly,
$$f = f p e r f t e q f = (f p e r)^n\, f\, (t e q f)^n = (f p e r)^{n+1}\, f\, (t e q f)^n = f p e r f.$$

We have $s = se = serfpe = srfpe$ and therefore, $[s][e]^\omega = [srfpe][erfpe]^\omega \subseteq L$. By strong recognition and since $[srfpe][erfpe]^\omega \cap [srf][fperf]^\omega \neq \emptyset$, we conclude $\beta \in [sr][f]^\omega = [srf][fperf]^\omega \subseteq L$. This shows that every infinite word in $L$ has an open environment contained in $L$. Every finite word $w$ has a trivial open environment $\{w\}$. Therefore $L$ is open. $\square$

**Lemma 4.7** *If $L \subseteq \Gamma^\infty$ is weakly recognized by $h: \Gamma^* \to M$ in **LDA** and if $L$ is closed in the strict $k$-factor topology for some $k \geq 2|M|$, then $L$ is definable in $\mathrm{FO}^2[<, +1]$.*

*Proof:* Let $\alpha \in L$ and $A = \mathrm{im}_k(\alpha)$. We can assume that $A = \{w_1, \ldots, w_s\} \neq \emptyset$ because $L \cap \Gamma^*$ is definable in $\mathrm{FO}^2[<, +1]$, see e.g. [14]. Write $\alpha = u \cdot w \cdot \beta$ with $w \notin \mathrm{alph}_k(\mathrm{last}_{k-1}(w) \cdot \beta)$ and $w$ is the last factor in $\alpha$ which occurs only finitely often. If all factors occur infinitely often, then we set $\alpha = \beta$. In the remainder, we assume that some factor appears finitely often; the other case is similar. Let $r$ be the Ramsey number for monochromatic triangles when using $|M|$ colors. We consider the following factorization of $\beta$:
$$\beta = u_1 v_1 u_2 \cdots u_{rs} v_{rs} \gamma$$

where $v_{i+1} = w_{(i \bmod s)+1}$ and $v_i \notin \mathrm{alph}_k(u_i \, \mathrm{first}_{k-1}(v_i))$, i.e., $v_{i+1}$ is always the first occurrence of this factor after $v_i$ and we iterate seeing all factors in $A$ for $r$-many times. We write $U_i$ for the set of words in $[h(u_i)] \cap \left((A \setminus \{v_i\})^{\circledast} \cup \Gamma^{<k}\right)$ which do not end with some non-empty suffix $x$, $|x| < k$, such that $v_i$ is a prefix of $xv_i$, i.e., no word in $U_i \cdot \mathrm{first}_{k-1}(v_i)$ admits $v_i$ as a factor. We define
$$P(\alpha) = [h(u)] \cdot w \cdot \left(A^{\circledast} \cap U_1 \, v_1 \, \cdots \, U_{rs} \, v_{rs} \circ A^{\circledast}\right) \cap A^{\text{\textcircled{im}}}$$

We have $\alpha \in P(\alpha)$. The remainder of the proof is divided into two parts. First, we show $P(\alpha) \subseteq L$ and second, we show that $P(\alpha)$ is definable by some formula $\varphi_\alpha \in \mathrm{FO}^2[<, +1]$ where the size of $\varphi_\alpha$ only depends on $M$ and $k$, but not on $\alpha$.

By choice of $r$, there exists $a \in M$ and an idempotent $e \in M$ such that every word $\alpha' \in P(\alpha)$ (including $\alpha$ itself) admits a factorization $\alpha' = u' \cdot w \cdot x' e'_1 e'_2 \beta'$ with $h(u') = h(u)$, $h(x') = a$, $h(e'_1) = h(e'_2) = e$, $\mathrm{alph}_k(e'_1) = \mathrm{alph}_k(e'_2) = \mathrm{alph}_k(x' e'_1 e'_2 \beta') = A = \mathrm{im}_k(x' e'_1 e'_2 \beta')$. For $\alpha$ we use the fixed factorization $\alpha = u \cdot w \cdot x e_1 e_2 \beta''$. Let now $\alpha' = u' \cdot w \cdot x' e'_1 e'_2 \beta' \in P(\alpha)$ be some arbitrary word in $P(\alpha)$. We want to show that $\alpha' \in \overline{L} = L$.

Let $z'$ be a finite prefix of $\beta'$. Let $z$ the suffix of $e'_2 z'$ of length $k$. By construction $z$ is a factor of $e_1$, i.e., $e_1 = y_1 z y_2$ for some $y_1, y_2 \in \Gamma^*$. Now, $x' e'_1 e'_2 z' \cdot y_2 e_2 \beta'' \in A^{\circledast} \cap A^{\text{\textcircled{im}}}$. We claim that $u' \cdot w \cdot x' e'_1 e'_2 z' \cdot y_2 e_2 \beta'' \in L$. To this end, it suffices to show $h(e'_2 \, z' \, y_2 \, e_2) = e$. We factorize $z' y_2 = x_0 \cdots x_m$ with $0 < |x_i| \leq |M|$ such that for every $i > 0$ the exists an idempotent $f_i \in h(\Gamma^+)$ such that $h(x_{i-1}) = h(x_{i-1}) f_i$. By construction and since $k \geq 2|M|$ we have $e \leq_{\mathcal{R}} h(x_0) f_1$, $e \leq_{\mathcal{L}} f_m h(x_m)$, $e \leq_{\mathcal{J}} f_i h(x_i) f_{i+1}$ (cf. proof of Lemma 4.6). Using Proposition 4.3, we conclude $h(e'_2 \, z' \, y_2 \, e_2) = e$.

Now, we show that $P(\alpha)$ is defined by some formula in $\mathrm{FO}^2[<, +1]$. In order to provide a concise notation, we introduce macros $\lambda(x) = w$ for a finite word $w$ expressing that the factor $w$ starts at position $x$; $\lambda(x) \in A$ for a finite collection of finite words $A$ as a shortcut for $\bigvee_{u \in A} \lambda(x) = u$; and finally $y > x + n$ and $y < x + n$ for $n \in \mathbb{N}$ with the natural interpretation. First, we verify that we see the sequence of $v_i$'s after the last factor $w$ and that after this last $w$ we do not have factors of length $k$ which are not in $A$. This is done by the formula

$$\exists x \colon \lambda(x) = w \land \forall y > x \colon \lambda(y) \in A \land \exists y \geq x + k \colon \nu_1(y) \tag{3}$$

with $\nu_i(x) \in \mathrm{FO}^2[<, +1]$ expressing that the suffix starting at $x$ is in $\Gamma^+ v_i \Gamma^* v_{i+1} \cdots \Gamma^* v_{rs} \Gamma^\infty$. This is achieved by the inductive construction $\nu_i(x) \equiv \exists y > x \colon \lambda(y) = v_i \land \exists x \geq y + k \colon \nu_{i+1}(x)$ for $i \leq rs$ and $\nu_i(x) \equiv \top$ else.

By the finite case [14], we see that $[h(u)]$ and every language $[h(u_i)]$ is definable in $\mathrm{FO}^2[<, +1]$ and hence so is $U_i$ because we can specify suffixes and words shorter than $k$ explicitely in $\mathrm{FO}^2[<, +1]$. Let $\mu, \mu_i \in \mathrm{FO}^2[<, +1]$ such that $[h(u)] = L(\mu)$ and $U_i = L(\mu_i)$. We use a relativization technique to restrict the interpretation of $\mu_i$ to the interval $I_i$ comprising all positions strictly between the $v_{i-1}$ and $v_i$ (with $v_0 = w$ for convenience).

For this we inductively construct formulas $\eta_i^<(x)$ and $\eta_i^>(x)$ in $\mathrm{FO}^2[<, +1]$

$$\eta_i^<(x) \equiv \exists y > x \forall x \leq y \colon \lambda(x) = v_i \Rightarrow \neg \eta_{i-1}^>$$
$$\eta_i^>(x) \equiv \exists y \colon x = y + k + 1 \land \neg \eta_i^<(y)$$

with $\eta_0^<(x) \equiv \exists y > x \colon \lambda(y) = w$ and $\eta_0^>(x) \equiv \exists y \colon x = y + k + 1 \land \neg \eta_0^<(y)$.

With these prerequisites we now define the relativization $\langle \psi \rangle_i$ of a formula $\psi$ to the interval $I_i$ by the rule $\langle \exists x \colon \xi \rangle_i \equiv \exists x \colon \eta_{i-1}^> \land \eta_i^< \land \langle \xi \rangle_i$. Boolean connectives and atomic variables are straightforward and the universal quantifier is then given by the equivalence $\forall x \colon \psi \equiv \neg \exists x \colon \neg \psi$.

Therefore, we get that $P(\alpha)$ is defined by the conjunction of the formula in (3) and the sentence $\langle \mu \rangle_0 \wedge \bigwedge_i \langle \mu_i \rangle_i$ in $\text{FO}^2[<, +1]$ where we set $\eta_{-1}^{\geq} = \top$ for brevity.

The size of this formula is bounded by a constant depending only on $|A|^k$ and $|M|$. These parameters do not depend on $\alpha$ and therefore, there are only finitely many languages $P(\alpha)$ when $\alpha$ varies over $L$. Now,
$$L = \bigcup_{\alpha \in L} P(\alpha)$$
is a finite union. Hence, $L$ is definable in $\text{FO}^2[<, +1]$. □

We are now ready to prove Theorem 4.1.

*Proof (Theorem 4.1):* "1 ⇒ 3" is Lemma 4.5. The implication "3 ⇒ 2" follows with Lemma 4.6; and "2 ⇒ 1" is Lemma 4.7. □

## 5 The first-order fragment $\Delta_2$

Over finite words, the fragments $\text{FO}^2[<, +1]$ and $\Delta_2[<, +1]$ have the same expressive power [14, 26]. This is not true for infinite words. Here, it turns out that $\Delta_2[<, +1]$ is a strict subclass of $\text{FO}^2[<, +1]$ and that the $\Delta_2[<, +1]$-languages are exactly the clopen languages in $\text{FO}^2[<, +1]$.

**Theorem 5.1** *Let $L \subseteq \Gamma^\infty$ be a language. The following are equivalent:*

(1) *$L$ is $\Delta_2[<, +1]$-definable.*

(2) *$L$ is $\text{FO}^2[<, +1]$-definable and clopen in the factor topology.*

*Proof:* "1 ⇒ 2": Since $L$ is definable in $\Delta_2[<, +1]$, we get that it is open by Theorem 3.1 and that it is closed by Theorem 3.4. Moreover, we get by these theorems that $\text{Synt}(L)$ is locally path-top as well as locally path-bottom. Proposition 4.3 yields $\text{Synt}(L) \in \mathbf{LDA}$ and Theorem 4.1 shows $L \in \text{FO}^2[<, +1]$.

"2 ⇒ 1": By Theorem 4.1 $\text{Synt}(L) \in \mathbf{LDA}$ and by Proposition 4.3 $\text{Synt}(L)$ is locally path-top as well as locally path-bottom. By Theorems 3.1 and 3.4 we get that $L$ is definable in $\Sigma_2[<, +1]$ and $\Pi_2[<, +1]$. □

A consequence of Theorem 5.1 is that $\Delta_2[<, +1]$ is a strict subclass of $\text{FO}^2[<, +1]$. In fact, it is a strict subclass of the intersection for the fragments $\text{FO}^2[<, +1]$ and $\Sigma_2[<, +1]$.

**Corollary 5.2** *Over $\Gamma^\infty$, the fragment $\Delta_2[<, +1]$ is a strict subclass of the fragment $\text{FO}^2[<, +1] \cap \Sigma_2[<, +1]$ and also of the fragment $\text{FO}^2[<, +1] \cap \Pi_2[<, +1]$.*

*Proof:* The set of non-empty finite words $\Gamma^+$ is defined by the sentence
$$\exists x \, \forall y \colon y \leq x$$
in $\text{FO}^2[<] \cap \Sigma_2[<]$. We have to show that $\Gamma^+$ is not definable in $\Pi_2[<, +1]$. By Theorem 3.4 it suffices to show that $\Gamma^+$ is not factor closed. Let $a \in \Gamma$, and consider the word $\alpha = a^\omega \notin \Gamma^+$. Every factor open set containing $\alpha$ also contains some finite word $a^m \in \Gamma^+$. Hence, the complement of $\Gamma^+$ is not factor open, and therefore, $\Gamma^+$ is not factor closed. By complementation, we see that $\Gamma^\omega$ is definable in $\text{FO}^2[<] \cap \Pi_2[<]$ but not in $\Delta_2[<, +1]$. □

**Example 5.3** We consider another language which is definable in $\mathrm{FO}^2[<] \cap \Sigma_2[<]$ but not in $\Delta_2[<,+1]$. Let $\Gamma = \{a,b\}$ and $L_3 = \Gamma^* ab^\infty$. The language $L_3$ is $\mathrm{FO}^2[<] \cap \Sigma_2[<]$-definable:

$$\exists x\, \forall y\colon \lambda(x) = a \;\wedge\; \bigl(\lambda(y) = a \;\Rightarrow\; y \leq x\bigr).$$

In order to show that $L_3$ is not definable in $\Pi_2[<,+1]$, it suffices to show that $L_3$ is not factor closed (Theorem 3.4). Let $k \in \mathbb{N}$. Every open set containing the word $(b^k a)^\omega \notin L_3$ also contains some word $(b^k a)^m b^\omega \in L_3$. Hence, the complement of $L_3$ is not $k$-factor open, and therefore, there is no $k$ such that $L_3$ is closed in the $k$-factor topology.

The same reasoning also works over $\Gamma^\omega$, since the language of all infinite words is definable in $\Pi_2[<,+1]$. Hence, $L_3' = \Gamma^* ab^\omega$ is definable in $\Sigma_2[<]$ over infinite words and in $\mathrm{FO}^2[<]$ but not in $\Delta_2[<,+1]$ over infinite words. The language $L_3'$ is the standard example of a language which cannot be recognized by a deterministic Büchi automaton [28, Example 4.2]. In particular, none of the fragments $\mathrm{FO}^2[<,+1]$ or $\Sigma_2[<,+1]$ contains only deterministic languages. ◇

**Example 5.4** Let $\Gamma = \{a,b,c\}$ and consider the language $L_4 = (\Gamma^2 \setminus \{bb\})^\circledast \circ aa \circ (\Gamma^2)^\circledast$ consisting of all words such that there is no factor $bb$ before the first factor $aa$. The language $L_4$ is defined by the $\Sigma_2[<,+1]$-sentence

$$\exists x \forall y < x\colon \lambda(x) = aa \wedge \lambda(y) \neq bb.$$

Here, $\lambda(x) = w$ is a shortcut saying that a factor $w$ starts at position $x$. A word $\alpha$ is in $L_4$ if and only if $aa$ is a factor of $\alpha$ and for every factor $bb$ there is a factor $aa$ to the left. These properties are $\Pi_2[<,+1]$-definable and hence $L_4 \in \Delta_2[<,+1]$. The language $L_4$ is not definable in any of the fragments $\mathrm{FO}^2[<]$, $\Sigma_2[<]$, or $\Pi_2[<]$ without successor, since its syntactic monoid is neither locally top nor locally bottom, cf. [7]. The language $L_4 \cap \Gamma^*$ has been used as an example of a language not definable in the Boolean closure of $\Sigma_2[<]$ over finite words by Almeida and Klíma [2, Proposition 6.1] as well as by Lodaya, Pandya, and Shah [14, Theorem 4]. The Boolean closure of $\Sigma_2[<]$ over finite words coincides with the second level of the Straubing-Thérien hierarchy, cf. [18, 27]. ◇

# 6 The first-order fragments $\mathrm{FO}^2 \cap \Sigma_2$ and $\mathrm{FO}^2 \cap \Pi_2$

In this section, we show that topological concepts can not only be used as an ingredient for characterizing first-order fragments, but also for describing some relations between fragments. More precisely, we relate languages definable in both $\Sigma_2[<,+1]$ and $\mathrm{FO}^2[<,+1]$ with the interiors of $\mathrm{FO}^2[<,+1]$-languages with respect to the factor topology. Dually, the languages in the fragment $\mathrm{FO}^2[<,+1] \cap \Pi_2[<,+1]$ are precisely the topological closures of $\mathrm{FO}^2[<,+1]$-languages. Remember that for a language $L$, its *closure* $\overline{L}$ is the intersection of all closed sets containing $L$. It can be "computed" as

$$\overline{L} = \{\alpha \in \Gamma^\infty \mid \forall U \subseteq \Gamma^\infty \text{ open with } \alpha \in U\colon U \cap L \neq \emptyset\}.$$

The *interior* of $L$ is the union of all open sets contained in $L$. The interior of a language is the complement of the closure of its complement.

**Theorem 6.1** *Let $L \subseteq \Gamma^\infty$ be a regular language. The following are equivalent:*

(1) $L \in \mathrm{FO}^2[<,+1] \cap \Sigma_2[<,+1]$.

(2) $L \in \mathrm{FO}^2[<,+1]$ *and $L$ is open in the factor topology.*

(3) $L$ *is the factor interior of some $\mathrm{FO}^2[<,+1]$-definable language.*

*Proof:* By complementation, the proof follows from Theorem 6.2 below. □

The equivalence of (1) and (2) is an immediate consequence of Theorems 3.1 and 4.1. The surprising property is (3); for example, it is not obvious that the factor interior of an $\mathrm{FO}^2[<,+1]$-definable language is again in $\mathrm{FO}^2[<,+1]$. It is slightly easier to first proof Theorem 6.2 — and then conclude Theorem 6.1 — than the other way round. The reason is that "computing" the closure is slightly easier than "computing" the interior.

**Theorem 6.2** *Let $L \subseteq \Gamma^\infty$ be a regular language. The following are equivalent:*

(1) $L \in \mathrm{FO}^2[<,+1] \cap \Pi_2[<,+1]$.

(2) $L \in \mathrm{FO}^2[<,+1]$ *and $L$ is closed in the factor topology.*

(3) $L$ *is the factor closure of some $\mathrm{FO}^2[<,+1]$-definable language.*

*Proof:* "1 ⇒ 2": If $L$ is in $\Pi_2[<,+1]$, then by Theorem 3.4, the language $L$ is factor closed.

"2 ⇒ 3": If $L$ is closed, then $\overline{L} = L$.

"3 ⇒ 1": By Theorem 4.1 and Theorem 3.4 it suffices to show that $\mathrm{Synt}(\overline{L})$ is in **LDA**. The factor interior of a regular language is regular. More precisely, a Büchi automaton recognizing the interior is effectively computable by Lemma 2.4. Since Büchi automata are effectively closed under complementation, the language $\overline{L}$ is regular. Let $\overline{L}$ be $k$-factor closed and let $n \geq |\mathrm{Synt}(L)| + |\mathrm{Synt}(\overline{L})| + k$, let $p \in \Gamma^+$, and let $q, r \in \Gamma^*$. We set

$$u = (p^n\, q\, p^n\, r\, p^n)^n\, p^n q p^n\, (p^n\, q\, p^n\, r\, p^n)^n,$$
$$v = (p^n\, q\, p^n\, r\, p^n)^n$$

We have

$$xuyz^\omega \in \overline{L} \iff xvyz^\omega \in \overline{L}$$

for all $x, y, z \in \Gamma^*$. By left-right symmetry, it suffices to show the implication from left to right. Let $s$ be a finite prefix of $z^\omega$. Since $xuyz^\omega \in \overline{L}$ there exists $\beta \in \mathrm{alph}_k(z^\omega)^\infty$ with $xuys \circ \beta \in L$. Then, since $\mathrm{Synt}(L) \in$ **LDA**, we have $xvys \circ \beta \in L$. Hence, $xvyz^\omega \in \overline{L}$. Moreover

$$x(uy)^\omega \in \overline{L} \iff x(vy)^\omega \in \overline{L}$$

for all $x, y \in \Gamma^*$. Again by left-right symmetry, it suffices to show the implication from left to right. Let $m \geq 1$ and consider the prefix $(vy)^m$ of $(vy)^\omega$. Since $x(uy)^\omega \in \overline{L}$ there exists $\beta \in \mathrm{alph}_k((uy)^\omega)$ with $x(uy)^m \circ \beta \in L$. By choice of $n$, we have $\mathrm{alph}_k((uy)^\omega) = \mathrm{alph}_k((vy)^\omega)$ and $\mathrm{last}_{k-1}((uy)^m) = \mathrm{last}_{k-1}((vy)^m)$. Since $\mathrm{Synt}(L) \in$ **LDA**, we have $x(vy)^m \circ \beta \in L$. Hence, $x(vy)^\omega \in \overline{L}$. □

## 7 Summary

We considered fragments of first-order logic over finite and infinite words. As binary predicates we allow order comparison $x < y$ and the successor predicate $x = y+1$. Figure 1 depicts the relation between the fragments $\Sigma_2[<,+1]$, $\Pi_2[<,+1]$, and $\mathrm{FO}^2[<,+1]$. Moreover, the languages $L_1, L_2, L_3$, and $L_4$ from Examples 4.4, 5.3, and 5.4 are included. For the other languages, we fix $\Gamma = \{a, b, c\}$ and $\emptyset \neq A \subsetneq \Gamma$.

The central notion for presenting our results is a partially defined composition $u \circ_k v = u'xv'$ where $u = u'x$, $v = xv'$, and $|x| = k-1$. Using this composition, one can show that the languages definable in $\Sigma_2[<,+1]$ is exactly the class of factor polynomials. Moreover, the composition $\circ_k$ leads to the $k$-factor topology, which we use in further characterizations of the successor fragments. A set is *factor open* if there exists some number $k$ such that $L$ is $k$-factor open. For every regular language $L$, Proposition 2.2 gives a bound $k$ such that $L$ is factor open if and only if $L$ is $k$-factor open. Then, in Proposition 2.3, we essentially show that for a given number $k$ it is decidable whether a regular language $L$ is $k$-factor open. Altogether, in order to check whether $L$ is factor open, we can check whether $L$ is $k$-factor open, with $k$ being the bound given by Proposition 2.2. Hence, the topological properties, which we use in the characterizations of the fragments, are decidable. Together with the decidable algebraic properties, this gives a decision procedure for deciding whether a given regular language $L \subseteq \Gamma^\infty$ or $L \subseteq \Gamma^\omega$ is definable in one of the fragments under consideration. In Table 1 we summarize our main results. All fragments are using binary predicates $[<,+1]$. The first decidable characterization of $FO^2[<,+1]$ is due to Wilke [31]. Decidability for $\Sigma_2[<,+1]$ over infinite words is new (Corollary 3.3).

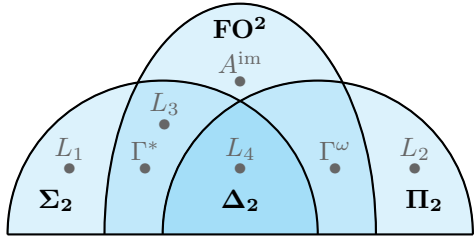

Figure 1: The fragments $\Sigma_2[<,+1]$, $\Pi_2[<,+1]$, and $FO^2[<,+1]$ over $\Gamma^\infty$.

| Logic | Languages | Algebra | + | Topology | |
|---|---|---|---|---|---|
| $\Sigma_2$ | factor polynomials | $eP_e e \leq e$ | + | factor open | Thm. 3.1 |
| $\Pi_2$ | | $eP_e e \geq e$ | + | factor closed | Thm. 3.4 |
| $FO^2$ | | **LDA** weak **LDA** | + | strictly factor closed | Thm. 4.1 |
| $\Delta_2$ | | **LDA** | + | factor clopen | Thm. 5.1 |
| $FO^2 \cap \Sigma_2$ | factor interior of $FO^2$ | **LDA** | + | factor open | Thm. 6.1 |
| $FO^2 \cap \Pi_2$ | factor closure of $FO^2$ | **LDA** | + | factor closed | Thm. 6.2 |

Table 1: Main characterizations of some first-order fragments

**Open problems** The fragment $\Sigma_2[<,+1]$ has a language description in terms of factor polynomials. Without the successor predicate similar characterizations in terms of so-called unambiguous polynomials exist for the fragments $FO^2[<]$, for $FO^2[<] \cap \Sigma_2[<]$, and for $\Delta_2[<]$, cf. [7]. It is open whether these fragments admit similar characterizations if we allow the successor predicate.

Moreover, for the fragment $\Delta_2[<,+1]$ we only have an implicit decidable characterization based on the decidability of $\Sigma_2[<,+1]$ and $\Pi_2[<,+1]$ (or alternatively, based on the decidability of $FO^2[<,+1]$ and being clopen). A more direct characterization of this fragment remains open. For $\Delta_2[<]$ without successor, such a characterization shows that all languages in $\Delta_2[<]$ over infinite words are recognizable by deterministic Büchi automata.

Another important fragment is $\mathbb{B}\Sigma_1$, the Boolean closure of $\Sigma_1$. A result of Knast [12] shows that, over finite words, it is decidable whether a regular language is definable in the logic $\mathbb{B}\Sigma_1[<,+1,\min,\max]$, which over finite words corresponds to the first level of the dot-depth hierarchy. A similar result over infinite words is still missing.

**Acknowledgments** We would like to thank the anonymous referees of the conference version of this paper for their comments which helped to improve the presentation.